\documentclass[%
 reprint,
 amsmath,amssymb,
 aps,
]{revtex4-2}

\usepackage{graphicx}
\usepackage{dcolumn}
\usepackage{bm}

\usepackage{graphicx,amsmath,amssymb,epsfig,epstopdf,color,float}
\usepackage{stfloats}
\usepackage{color}
\usepackage{lineno}
\usepackage{enumerate}
\usepackage{comment}\allowdisplaybreaks

\makeatletter
\newcommand*{\centerfloat}{%
  \parindent \z@
  \leftskip \z@ \@plus 1fil \@minus \marginparwidth
  \rightskip \leftskip
  \parfillskip \z@skip}
\makeatother

\begin{document}


\title{Quantum reflection of single photons in a cold Rydberg atomic gas}
\thanks{The article has been accepted and published by Optics Letters.}%

\author{Yao Ou$^{1}$, Qi Zhang$^{1}$, and Guoxiang Huang$^{1,2,3}$
       }
       \homepage{gxhuang@phy.ecnu.edu.cn}
\affiliation{$^1$State Key Laboratory of Precision Spectroscopy,
                 East China Normal University, Shanghai 200062, China\\
             $^2$NYU-ECNU Joint Institute of Physics, New York University at Shanghai, Shanghai 200062, China\\
             $^3$Collaborative Innovation Center of Extreme Optics, Shanxi University, Taiyuan 030006, China
             }

\date{\today}

\begin{abstract}
We propose and analyze a scheme for realizing the quantum reflection of single photons in a cold Rydberg atomic gas via electromagnetically induced transparency, by which a deep and tunable attractive potential well can be prepared by using stored gate photons. Such a scheme is promising for designing dispersion-type single-photon switches, and may be taken as a  quantum device for observing the wave and particle natures of photons simultaneously.
\end{abstract}

\maketitle


\section{\label{sec:level1}Introduction}

Quantum reflection (QR) refers to a reflection of quantum particles from an external potential with unit probability even in the absence of a classically turning point~\cite{Friedrich2004}. Such a reflection looks paradoxical because it is very counterintuitive and nonclassical. Typical example of QR is the one when a quantum particle is reflected from an attractive potential well. QR manifests the wave nature of quantum particles and cannot be described by Wentzel, Kramers and Brillouin (WKB) theory~\cite{Petersen2018,BaiZY2020PRA}.
In recent years, much attention has been paid to the study on QR in various systems due to their intriguing physics and practical applications~\cite{Friedrich2004,Petersen2018,BaiZY2020PRA}.

On the other hand, dual wave–particle (WP) nature of light is a research topic with a very long history. After centuries of debate, it was finally confirmed that light is not only a wave but also a particle. WP duality has been recognized to be the most basic nature of all quantum particles and played a central role in the establishment of quantum mechanics. Bohr explained the WP duality with his well-known complementarity principle, which states that
a quantum object behaves as a wave {\it or} as a particle (which are mutually exclusive), depending on detecting devices~\cite{Bohr1928}. However, many investigations indicated that the Bohr's complementarity principle is inadequate. Especially, simultaneous observation of (or morphing between) the wave and particle aspects of photons were reported in recent
years~\cite{Yan2015,Ma2016,
Wang2016,Rab2017,Li2017,Qian2018,Yoon2021,Siddiqui2021}.

In this Letter, we propose a scheme to realize the QR of single photons in a cold Rydberg atomic gas via electromagnetically induced transparency (EIT). Based on the strong Rydberg-Rydberg interaction, we show that a deep and actively tunable attractive (Rydberg defect) potential well can be obtained by storing gate photons, and hence the QR of a single probe photon can be made very efficient, which can be used to design single-photon switch. Such a switch is of {\it dispersion-type} and hence possesses excellent quantum coherence; it is different from those suggested recently (see Ref.~\cite{Murray2016NewYork} and references therein) which are of dissipation-type in nature. Moreover, the QR of single photons proposed here provides a new route for simultaneously observing the wave and particle natures of photons since the incident probe laser field is in a single-photon state and its reflection by the attractive potential well is a pure wave behavior.

\begin{figure*}[htbp]
\centering
\fbox{\includegraphics[width=0.8\linewidth]{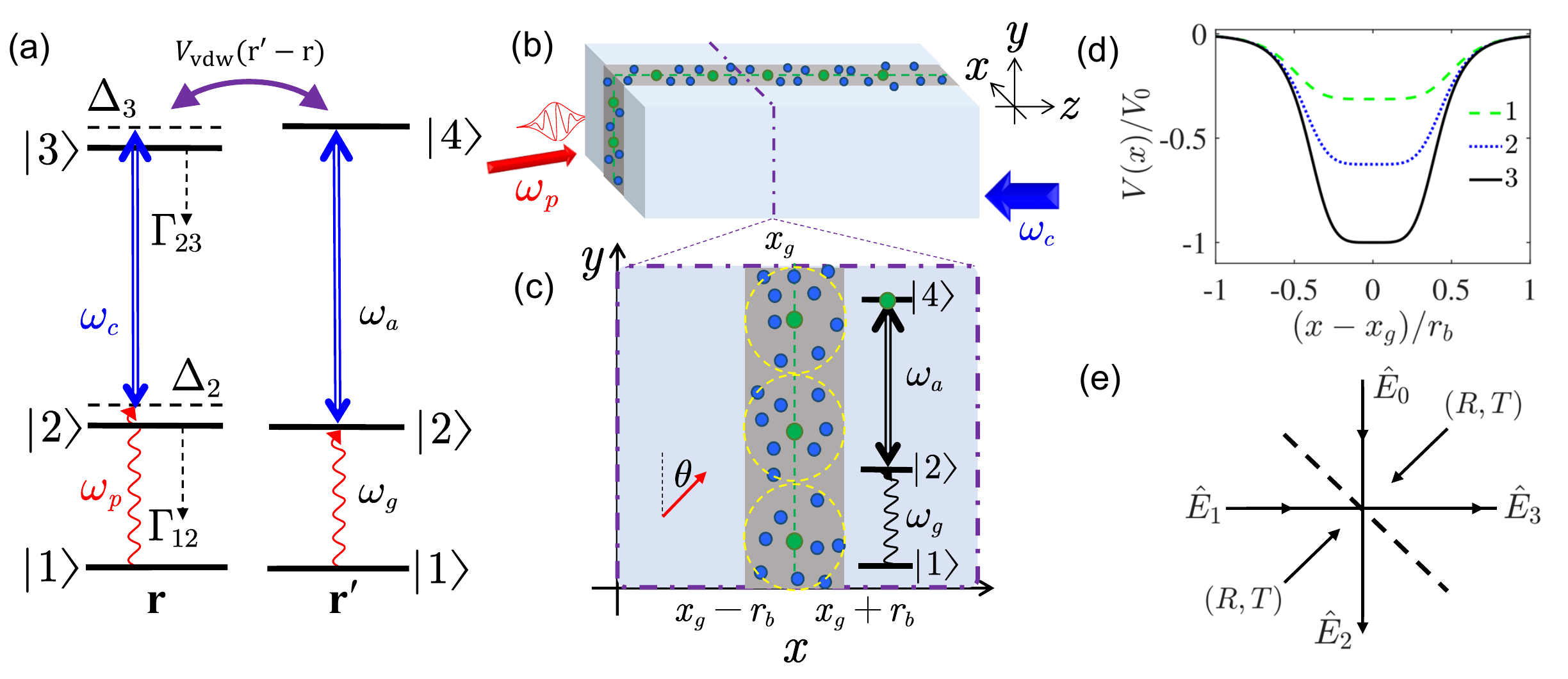}}
\caption{\footnotesize
(a)~Level diagram and excitation scheme of the Rydberg-EIT.
(b)~Geometry for observing the probe-photon scattering from the Rydberg defect.
(c)~Rydberg defect (prepared via another Rydberg-EIT) located at the plane $x=x_g$, realized by storing gate photons in the Rydberg state $|4\rangle$ of the gate atoms. The level diagram drawn on the left part is the same as that of the right part of (a); $\theta$ is the incident angle of the probe photon. For more details, see text.
(d)~Shape of the dimensionless Rydberg defect potential  $V(x)/V_0$ for $^{88}$Sr atomic gas
as a function of $(x-x_g)/r_b$ ($r_b$ is Rydberg blockade radius). Lines 1, 2 and 3 are for  $\Delta_{2}=-2\pi\times320$MHz, $-2\pi\times160$MHz, and $-2\pi\times100$MHz, respectively. 
(e)~Beam-splitter description for the single-photon incidence onto to Rydberg attraction potential well. The black-dashed line denotes the position of the potential well. $\hat{E}_1$ represents the incident probe field; $\hat{E}_2$ and $\hat{E}_3$ are respectively reflected and transmitted fields; $\hat{E}_0$ is a vacant incident field, required by a complete quantum description of beam splitter. $\sqrt{R}$ and $\sqrt{T}$ are the reflection and transmission amplitudes, respectively.}
\label{Fig1}
\end{figure*}

\section{Model and Rydberg-defect potential}
We consider a cold atomic gas under the excitation of
Rydberg-EIT~\cite{Pritchard2010PRL}; see the left part of Fig.~\ref{Fig1}(a).

A weak, quantized probe laser field $\hat{\bf E}_p$ (angular frequency $\omega_{p}$; propagating along $z$ direction) couples the ground state $|1\rangle$ and the intermediate state $|2\rangle$; a strong control field $\bf E_c$ (angular frequency $\omega_{c}$; propagating along $-z$ direction) couples $|2\rangle$ and the Rydberg state $|3\rangle$ (with principle quantum number $n$; $n=60$ is chosen for the numerical calculations given below); $\Delta_2$ and $\Delta_3$ are respectively one- and two-photon detunings; $\Gamma_{12}$ and $\Gamma_{23}$ are respectively the spontaneous emission decay rates from $|2\rangle$ to $|1\rangle$ and $|3\rangle$ to $|2\rangle$. Fig.~\ref{Fig1}(b) is a schematic of the geometry for observing the QR of incident probe photons.

To prepare a Rydberg-defect potential, we assume that gate photons are stored in another Rydberg state $|4\rangle$ of the gate atoms (principal quantum number $n'=60$) located in an array of at $x=x_g$, realized by using two laser fields (angular frequencies $\omega_{g}$ and $\omega_{a}$ respectively); see the right part of Fig.~\ref{Fig1}(c) [also Fig.~\ref{Fig1}(a)]~\cite{BaiZY2020PRA}. Because $|3\rangle$ and $|4\rangle$ are Rydberg states,
the atoms between such states have a strong interaction described by the van-der-Waals (vdW) potential $V_{\rm vdW}=-\hbar C_6/|{\mathbf r}_g-{\bf r}|^6$, where ${\bf r}_g=(x_g, y, z)$ and $C_6$ is dispersion parameter. In Fig.~\ref{Fig1}(c), the region of the Rydberg defect is illustrated by the grey domain, where dashed yellow circles are Rydberg-blockade spheres of radius $r_b$ and green (blue) dots are
atoms in the Rydberg state $|4\rangle$ (ground state $|1\rangle$).

The Hamiltonian of the system reads ${\hat H}={\hat H}_{\rm AF}+{\hat H}_{\rm AG}$, here ${\hat H}_{\rm AF}$  (${\hat H}_{\rm AG}$) describes the atom-light interaction (Rydberg-Rydberg interaction between the atoms in states $|3\rangle$ and $|4\rangle$). For explicit expressions of the electric field $\hat{\bf E}={\hat{\bf E}}_{p}+{\bf E}_{c}$ and the Hamiltonian ${\hat H}$, see Sec.~S1\,A of {\color{blue} Supplement 1}. Under electric-dipole, rotating-wave, and slowly-varying envelope approximations,
the dynamics of the atoms and the light field obey the Heisenberg-Langevin-Maxwell (HLM) equations~\cite{Bai2019}
\begin{subequations}
\begin{eqnarray}\label{HLME}
&& i\hbar \frac{\partial {\hat S}_{\alpha\beta}}{\partial t}=\left[{\hat S}_{\alpha\beta},{\hat H}\right]+\Gamma\,[{\hat S}_{\alpha\beta}]+{\hat F}_{\alpha\beta},\label{HLME1}\\
&& i\left(\frac{\partial}{\partial z}+\frac{1}{c}\frac{\partial}{\partial t}\right){\hat E}_{p}+\frac{1}{2k_{p}}\nabla_{\perp}^2{\hat E}_{p}+\frac{g_{p}^{\ast}N}{c}{\hat S}_{21}=0.\label{HLME2}
\end{eqnarray}
\end{subequations}
Here, $\hat{S}_{\alpha\beta}$ are atomic transition operators associated with the states $|\alpha\rangle$ and $|\beta\rangle$ $(\alpha,\beta = 1,2,3,4$);
$\Gamma$ is a relaxation matrix describing the spontaneous emission and dephasing; ${\hat F}_{\alpha\beta}$ are $\delta$-correlated Langevin noise operators; $\nabla_{\perp}^2=\partial^2/\partial x^2+\partial^2/\partial y^2$; $g_{p}$ is single-photon half Rabi frequency; $N$ is the total atomic number; $\hat{E}_p$ is the annihilation operator of probe photons, satisfying  commutation relation $[\hat{E}_p({\bf r},t), \hat{E}_p^{\dag}({\bf r}',t)]=V \delta ({\bf r}-{\bf r}')$, with $V$ the quantized volume.

We assume that the all atoms (except for those in the Rydberg defect region) are prepared initially in the ground state $|1\rangle$, and the incident probe field  is at single-photon level and has a larger time duration, so that it can be taken as a weak continuous wave. Thereby, Eq.~(\ref{HLME1}) can be linearized  and the solution of ${\hat S}_{21}$ can be obtained easily. Then, from Eq.~(\ref{HLME2}) we obtain
\begin{align}\label{SE}
& i\hbar\frac{\partial}{\partial\tau}{\hat E}_{p}({\bf r}_{\perp},\tau)=
\left[-\frac{\hbar^2}{2m_{p}}\nabla_{\perp}^2+V(x)\right]{\hat E}_{p}({\bf r}_{\perp},\tau),
\end{align}
where $\tau= z/c$;  $m_{p}=\hbar k_{p}/c$;
$ V(x)\equiv -(\hbar \omega_{p}/2)\chi_{p}^{(1)}(x)$ is a potential originated from the Rydberg defect, with the linear optical susceptibility given by
$\chi_{p}^{(1)}(x)=\frac{{\cal N}_{a}|{\bf e}_{p}\cdot{\bf p}_{21}|^2}{\varepsilon_{0}\hbar}\frac{d_{31}-\Delta_{d}(x)}{|\Omega_{c}|^2
-d_{21}[d_{31}-\Delta_{d}(x)]}$.
Here, $d_{\alpha\beta}=\Delta_{\alpha}-\Delta_{\beta}+i\gamma_{\alpha\beta}$;
$\Omega_c\equiv(\mathbf{e}_c\cdot \mathbf{p}_{32})\mathcal{E}_c/\hbar$ is the half Rabi frequency of the control field; $\mathbf{p}_{\alpha\beta}$ is the electric dipole matrix elements associated with the states $|\beta\rangle$ and $|\alpha\rangle$; the $x$-dependent detuning, defined by $\Delta_{d}(x)\equiv - \pi {\cal N}_{g}{\cal N}_{x}^{-1} C_{6}/[2\left|x-x_{g}\right|^{4}]$, is contributed by the gate photons stored in the Rydberg state $|4\rangle$. Explicit expressions of Eqs.~(\ref{HLME1}) and (\ref{HLME2}), and
the derivation of Eq.~(\ref{SE})
are presented in Sec.~S1\,A-C of  the {\color{blue} Supplement 1}.

Generally, the Rydberg-defect potential $ V(x)$ has real and the imaginary parts. To make the imaginary part much smaller than the real part,
so that the absorption of the probe photon can be substantially inhibited, we assume that the system work in a {\it dispersion regime}. This can be realized by using the conditions $|\Omega_c|^2\gg \gamma_{21}\gamma_{31}$ and $|\Delta_{2}|\gg\gamma_{21}$. Moreover, due to the highly-manipulated character of the system, $V(x)$ can be tuned by adjusting the systemic parameters; especially, one can obtain an attractive (repulsive) potential by taking a negative (positive) one-photon detuning $\Delta_{2}$. This point is discussed in detail in Sec.~S2 of the {\color{blue} Supplement 1}. Fig.~\ref{Fig1}(d) shows the profile of dimensionless Rydberg defect potential $V(x)/V_0$ for $^{88}$Sr atomic gas\, ($n=n'=60$; $C_{6}= 2 \pi \times 81.6 \mathrm{GHz} \mu \mathrm{m}^{6}$) as a function of $(x-x_g)/r_b$. Here $V_0= \frac{\hbar^{2}}{2m_{p}R_{0}^{2}}$, $r_b= [|C_6 d_{21}|/(2|\Omega_c|^2)]^{1/6}$ is Rydberg blockade radius.
Lines 1, 2 and 3 are for  $\Delta_{2}=-2\pi\times320$MHz, $-2\pi\times160$MHz, and $-2\pi\times100$MHz, respectively. The other parameters used for plotting the figure are $\Delta_3=0$, $\Gamma_{12}=2\pi\times16\,{\rm MHz}$, $\Gamma_{23}=2\pi\times16.7\,{\rm kHz}$, ${\cal N}_{a}=3\times10^{10}\,{\rm cm}^{-3}$, $\Omega_c=2\pi\times16$MHz, $r_b=5.4\mu$m, and hence  $V_0\approx5.5 \times 10^{-23} $J. One sees that a deep, tunable Rydberg-defect potential well can indeed be realized in the system.

\section{Quantum reflections of single photons}
Equation (\ref{SE}) can be written in the form of Heisenberg equation of motion
\begin{align}\label{HE_eff}
& i\hbar\frac{\partial}{\partial\tau}{\hat E}_{p}({\bf r}_{\perp},\tau)=\left[{\hat E}_{p}({\bf r}_{\perp},\tau),{\hat H}_{\rm eff}^{\rm H}\right].
\end{align}
Here ${\hat H}_{\rm eff}^{\rm H}\equiv\frac{1}{S_{0}}\int d^2r_{\perp}{\hat E}_{p}^{\dag}({\bf r}_{\perp},\tau)\left[-\frac{\hbar^2}{2m_{p}} \nabla_{\perp}^2+V(x)\right]{\hat E}_{p}({\bf r}_{\perp},\tau)$ is an effective Hamiltonian, with ${\bf r}_{\perp}\equiv(x,y)$,  $d^2r_{\perp}\equiv dxdy$, and $m_p$ the effective mass of probe photons;
${\hat E}_{p}({\bf r}_{\perp},\tau)$ satisfies the commutation relation $[{\hat E}_{p}({\bf r}_{\perp},\tau), {\hat E}_{p}^{\dag}({\bf r}_{\perp}',\tau)]=S_{0}\delta({\bf r}_{\perp}'-{\bf r}_{\perp})$, with $S_{0}$ the quantization area of the probe field in the $x$-$y$ plane.

Since our aim is to study the scattering of single photons by the Rydberg defect potential, for which the dynamical information of a single-photon state is required, it is more convenient to convert the problem into a Schr\"{o}dinger picture. The state vector of the system, $|\Psi(\tau)\rangle$, satisfies the Schr\"{o}dinger equation $i\hbar\frac{\partial}{\partial\tau}|\Psi(\tau)\rangle={\hat H}_{\rm eff}^{\rm S}|\Psi(\tau)\rangle$, where ${\hat H}_{\rm eff}^{\rm S}$ is the effective Hamiltonian in the Schr\"{o}dinger picture.

We assume that the input probe field is in a one-photon state, then at "time" $\tau$ this state will become
$|\Psi(\tau)\rangle=\frac{1}{\sqrt{S_{0}}}\int d^2r_{\perp}\Phi({\bf r}_{\perp},\tau){\hat E}_{p}^{\dag}({\bf r}_{\perp})|0\rangle$.
Here $\Phi({\bf r}_{\perp},\tau)=\langle0|{\hat E}_{p}({\bf r}_{\perp})|\Psi(\tau)\rangle/\sqrt{S_{0}}$ is one-photon wavefunction (with $|\Phi({\bf r}_{\perp},\tau)|^2$ the probability for observing the photon at position ${\bf r}_{\perp}$ and time $\tau$, obeying $\int d^2r_{\perp}|\Phi({\bf r}_{\perp},\tau)|^2=1$). Then, it is easy to obtain the equation
\begin{equation}\label{dSP}
i\frac{\partial}{\partial\tau^{\prime}}\Phi=\left[-\left(\frac{\partial^2}{\partial \xi^2}+\frac{\partial^2}{\partial \eta^2}\right)+\tilde{V}(\xi)\right]\Phi.
\end{equation}
Here, $(\xi, \eta)= (x,y)/R_0$ and $\tau^{\prime}=\tau/\tau_{0}$\,
($\tau_{0}=\hbar/[2 m_{p} R_{0}^{2}]$; $R_{0}$ is the transverse size of the probe beam) are dimensionless variables, ${\tilde V}(\xi)\equiv V/V_0$ is dimensionless Rydberg-defect potential. For a detailed derivation of Eq.~(\ref{dSP}), see Sec.~S3 of the {\color{blue} Supplement 1}. Because the system works in the dispersion regime of the Rydberg-EIT, the evolution of the system is approximately unitary, and the single-photon character of the probe field can keep well during evolution. Thus the small imaginary part
of the potential ${\tilde V}$ in Eq.~(\ref{dSP}) can be neglected safely.

We now consider the scattering behavior of probe photons beyond WKB theory.
We assume a probe photon is incident onto the Rydberg-defect potential $\tilde{V}(\xi)$ from the left-hand side (with the $y$-$z$ plane as incident plane), so the reflection and transmission of the incident photon can be investigated by using Eq.~(\ref{dSP}) by taking $\partial/\partial \eta=0$. We find that reflection, transmission, or trapping of the photon can occur generally. These scattering behaviors can be described by coefficients of reflection ($R$),
transmission ($T$), and trapping ($L$),
defined by~\cite{Khawaja}
\begin{subequations}
\begin{eqnarray}
&& R=\frac{\int_{-\infty}^{\xi_{l}} d \xi\left|\Phi\left(\xi, \tau^{\prime}=L_{\mathrm{m}}\right)\right|^{2}}{\int_{-\infty}^{+\infty} d \xi|\Phi(\xi, \tau^{\prime}=0)|^{2}},\\
&& T=\frac{\int_{\xi_{r}}^{+\infty} d \xi\left|\Phi\left(\xi, \tau^{\prime}=L_{\mathrm{m}}\right)\right|^{2}}{\int_{-\infty}^{+\infty} d \xi|\Phi(\xi, \tau^{\prime}=0)|^{2}}, \\
&& L=\frac{\int_{\xi_{l}}^{\xi_{r}} d \xi\left|\Phi\left(\xi, \tau^{\prime}=L_{\mathrm{m}}\right)\right|^{2}}{\int_{-\infty}^{+\infty} d \xi|\Phi(\xi, \tau^{\prime}=0)|^{2}},
\end{eqnarray}
\end{subequations}
respectively.  Here $L_{m}$ is the length of the medium along $\tau^{\prime}$ direction; $\xi_{l}$ ($\xi_{r}$) is the position on the $\xi$ axis, at which the influence of the potential on the left-hand (right-hand) side of the defect potential is negligible. One has $T + L + R \approx 1$ because the evolution of the system is approximately unitary.
In our simulation below, Gaussian wavepacket $\Phi(\xi,0)=A_0\,e^{-(\xi+8)^2/18}\,e^{iv_{0}\xi}$ will be used, with $A_0$ and $v_0$ are dimensionless amplitude and incident velocity.

\begin{figure*}
\centering
\includegraphics[width=0.92\linewidth]{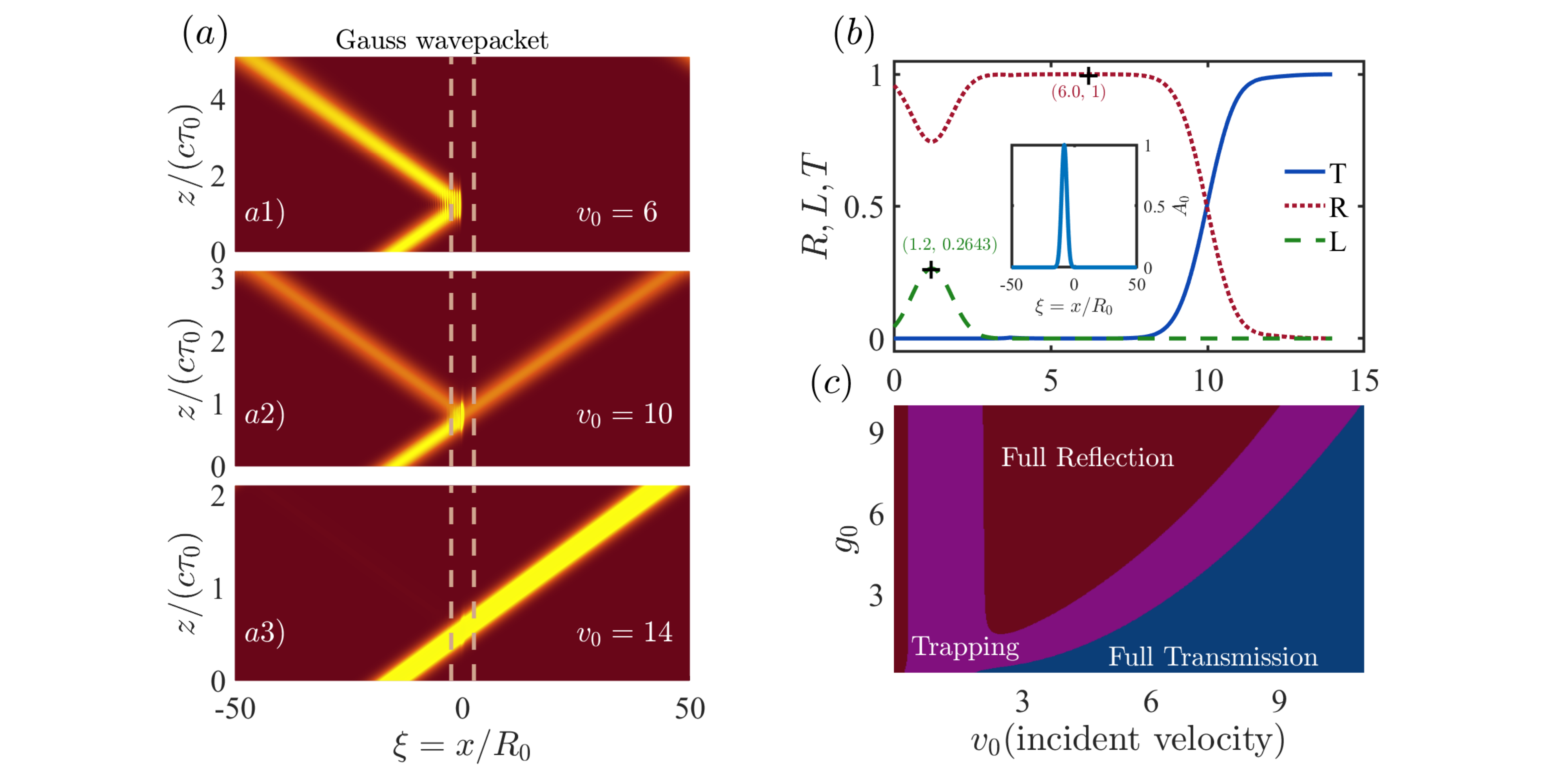}  
\caption{{\footnotesize Quantum reflection of probe photons.
(a)~Scattering of a single-photon Gaussian wavepacket in the plane of $\xi=x/R_0$ and $z/(c\tau_0)$, with the dimensionless incident velocity (a1)~$v_{0}=6$ (complete reflection), (a2)~$v_{0}=10$ (partial reflection and transmission), and
(a3)~$v_{0}=14$ (complete transmission), respectively.
The width of the Rydberg defect potential well is marked by two dashed grey lines.
(b)~Reflection coefficient $R$ (dotted red line), transmission coefficient $T$ (solid blue line) and trapping coefficient $L$ (dashed green line) of the single-photon wavepacket as functions of $v_{0}$.
$R$ reaches its maximum value 1 (full reflection) at $v_0=6.0$ (the ``+'' point at the dotted red line); $L$ reaches its maximum value 0.2643 at $v_0=1.2$ (the ``+'' point at the dashed green line). The inset shows the profile of the incident  single-photon Gaussian wavepacket (with width of $R_{0}=6\mu$m and maximum amplitude $A_{0}=1$).
(c) ``Phase diagram'' of the single photon scattering by taking $R$ as a function of $v_{0}$ and $g_{0}$ (the depth parameter of the Rydberg defect potential; see the text for detail).
The domains with dark red, deep blue, and purple colors are ones for full reflection, full transmission, and trapping of the single-photon wavepacket, respectively.
}}
\label{Fig2}
\end{figure*}
Fig.~\ref{Fig2}(a) shows the numerical result of the scattering of a single-photon wavepacket in the plane of $\xi=x/R_0$ and $z/(c\tau_0)$ for Rydberg blockade radius $r_{b}=5.4\mu$m. The width of the Rydberg defect potential well is marked by two dashed grey lines. Plotted in panels (a1), (a2), and (a3) are complete reflection (occurring $v_{0}=6$),
partial reflection and transmission (for $v_{0}=10$), and complete transmission (for $v_{0}=14$), respectively. The relation between the incident velocity $v_0$ and the incident angle $\theta$ [see also Fig.~\ref{Fig2}(c)]
is given by $\theta=\tan ^{-1}\left[v_0 /\left(R_{0}\right)\right]$; $R_{0}=6\mu {\rm m}$ in our simulation.

The scattering coefficients $R$, $T$, and $L$ as functions of $v_{0}$ are shown in Fig.~\ref{Fig2}(b) by the dotted red, solid blue, and dashed green lines, respectively. One sees that $R$ reaches its maximum value 1 (i.e. full reflection) at $v_0=6.0$, indicated by the ``+'' point at the dotted red line. Given in the inset of the figure is the profile is the incident single-photon Gaussian wavepacket, with width of $R_{0}=6\mu$m and maximum amplitude $A_{0}=1$. Due to attractive character of the Rydberg defect, a trapping of the wavepacket during the scattering process can also occur; $L$ reaches its maximum value 0.2643 at $v_0=1.2$, which is indicated by the ``+'' point at the dashed green line.

Fig.~\ref{Fig2}(c) shows the ``Phase diagram'' of the photon scattering, by taking reflection coefficient $R$ as a function of $v_{0}$, and $g_{0}$ [the depth parameter of the Rydberg defect potential, defined by (S24) of the {\color{blue} Supplement 1}].
In the figure, the domains with dark red, deep blue, and purple colors are ones for full reflection, full transmission, and trapping (crossover region from reflection to transmission) of the single-photon wavepacket. We stress that the
QR behaviors predicted above can also be observed by using other types of incident
singe-photon wavepackets; see the discussion given in Sec.~S4 of the {\color{blue} Supplement 1}.

Form these results, we see that the single-photon wavepacket can display various scattering behaviors; in particular, it can be totally reflected by the Rydberg defect. Such a QR is not possible for classical particles, for which no reflection can happen when they are incident to an attractive potential.
Physically, the QR can be understood as a specific interference phenomenon of incoming and outgoing waves when single photon interacts with the attractive Rydberg-defect potential, absent for the scattering of classical particles because such a phenomenon cannot be obtained based on Newtonian mechanics.

\section{Realizing dispersion-type single photon switches}
Single-photon switches are devices by which the transmission of a target photon can be controlled by a gate photon without the detour via electronic signal processing. In recent years, significant efforts have been made toward to realize such switches, as they lie at the heart of many practical applications of quantum information and precision measurement.

The effective QR of probe photons here can be applied to realize single-photon switches. In this protocol, the incident photon is the target photon and the Rydberg defect potential plays the role of a gate photon (i.e.
the photon stored in the Rydberg state $|4\rangle$). The preparation of gate photons is now a technique easily to be implemented, and a deep attractive Rydberg defect potential can also be realized by current-day experiments. Through controlling one-photon detuning $\Delta_2$~\cite{First2013} or selecting atomic species with different $C_6$~\cite{BaiZY2020PRA}, the incident target photon can be completely reflected [Fig.~\ref{Fig2}(a)] or trapped (see the Sec.~S5 of the {\color{blue} Supplement 1})
by the stored gate photon. We stress that the single-photon switch proposed here is of {\it dispersion-type} character and hence possesses excellent quantum coherence. It is different from those suggested recently (see Ref.~\cite{Murray2016NewYork} for details) which are photon switches of dissipation-type and hence can only be applied to classical domain in principle~\cite{Murray2016NewYork}.

\section{Testing the WP duality of photons}

Classical wisdom of WP duality (by Bohr et al.) regulates that a quantum object displays {\it either} particle {\it or} wave nature, but never both. Based on this viewpoint, it would be impossible to observe wave and particle natures of quantum particles simultaneously. Our study provides a way to test the WP duality of photons, which can be implemented by measuring the QR of the incident probe photon from the attractive Rydberg defect potential well, prepared by gate photons. In our protocol, the wave and particle natures of the photon can be detected simultaneously because the incident photon is in a single-particle quantum state and the QR is a pure wave behavior absent for classical particles.

The model suggested in Fig.~\ref{Fig1}(a) and (b) can be abstracted as a special beam splitter (BS), shown in Fig.~\ref{Fig1}(e). In the figure, the black-dashed line denotes the position of the Rydberg defect potential well; $\hat{E}_1$ denotes the light field of the incident probe photon; $\hat{E}_2$ and $\hat{E}_3$ are respectively the reflected and transmitted fields; $\hat{E}_0$ is a vacant incident field, required by a complete quantum description of the BS (for satisfying Heisenberg commutation relation). Since the system works under the condition of EIT, the loss of the BS is negligible. We assume that the Rydberg defect potential is not too deep, hence the trapping of the incident photon plays no significant role for large incident velocity. By suitably selecting system parameters ($g_{0}=10$, $\Delta_{2}=-2\pi\times160$MHz, $v_0=10$), we can obtain the relations $\hat{E}_2\approx -i \sqrt{T} \hat{E}_0+\sqrt{R} \hat{E}_1$, $\hat{E}_3\approx i \sqrt{R} \hat{E}_0+\sqrt{T} \hat{E}_1$ [see Fig.~\ref{Fig2}(b)]. Moreover,
the system can behave like a 50:50 BS when $T=R=0.5$. Due to the quantum behavior  of the incident photon, it is impossible to register the photon at the same time by the two photon counters placed respectively at the reflection and transmission ports (i.e. no coincidence-counting is expected)~\cite{Gerry2005}. Different from conventional BS, the observed zero coincidence-counting rate by using such a BS can manifest simultaneously both the particle and wave natures of the probe photon because the QR is a pure wave phenomenon.

\section{Conclusion}

We have proposed and analyzed a scheme for realizing the QR of single photons in a Rydberg atomic gas via EIT, where an active tunable, attractive Rydberg defect potential well is prepared by storing gate photons. We have shown that such a setting is useful not only for realizing dispersion-type single-photon switches, but also for designing  quantum devices to observe the particle and wave natures of photons simultaneously. The scheme can be extended to investigate multi-photon QRs and entanglements and find promising applications.

\begin{acknowledgments}

The project was funded by the National Natural Science Foundation of China (11975098). Data underlying the results presented in this Letter are herein and/or in {\color{blue}Supplement 1}, which can be obtained from the author.

\end{acknowledgments}


\end{document}